\def\authorone{Jelle F. Sleeboom}
\def\authortwo{Panayiotis Voudouris}
\def\authorthree{Melle T.J.J.M. Punter}
\def\authorfour{Frank J Aangenendt}
\def\authorfive{Daniel Florea}
\def\authorsix{Paul van der Schoot}
\def\authorlast{Hans M. Wyss}
\def\tueicms{Institute for Complex Molecular Systems, Eindhoven University of Technology, Eindhoven, the Netherlands}
\def\tuemech{Department of Mechanical Engineering, Materials Technology, Eindhoven University of Technology, Eindhoven, the Netherlands}
\def\amolf{AMOLF, Theory Biomol. Matter, Science Park 104, NL-1098 XG Amsterdam, the Netherlands}
\def\utrechtphys{Department of Physics, Utrecht University, Utrecht, the Netherlands}
\def\tuephys{Department of Physics, Eindhoven University of Technology, Eindhoven, the Netherlands}
\def\myabstract{We use dedicated microfluidic devices to expose soft hydrogel particles to a rapid change in the externally applied osmotic pressure and observe a non-monotonic response: After an initial rapid compression the particle slowly reswells to approximately its original size. Using a simple phenomenological and a more elaborate poroelastic model, we extract important material properties from a single microfluidic experiment, including the compressive modulus, the gel permeability and the diffusivity of the osmolyte inside the gel. We expect our approach to be relevant to applications such as controlled release, chromatography, and responsive materials.}

\documentclass[prl,twocolumn,showpacs,superscriptaddress,floatfix]{revtex4}

\usepackage{amsmath,latexsym}
\usepackage{float}
\usepackage{caption}
\usepackage{epsfig}

\usepackage{color}

\usepackage[normalem]{ulem}
\useunder{\uline}{\ul}{}

\begin{document}
\title{Compression and reswelling of microgel particles after an osmotic shock}

\author{\authorone}  
\thanks{These two authors contributed equally to this work.}
\affiliation{\tueicms}
\affiliation{\tuemech}
\author{\authortwo}  
\thanks{These two authors contributed equally to this work.}
\affiliation{\tueicms}
\affiliation{\tuemech}
\author{\authorthree} 
\affiliation{\amolf}
\author{\authorfour} 
\affiliation{\tueicms}
\affiliation{\tuemech}
\author{\authorfive} 
\affiliation{\tueicms}
\affiliation{\tuemech}
\author{\authorsix} 
\affiliation{\tuephys}
\affiliation{\utrechtphys}
\author{\authorlast} 
\email{H.M.Wyss@tue.nl}
\affiliation{\tueicms}
\affiliation{\tuemech}
\begin{abstract}
\noindent
\myabstract

\end{abstract}
\date{\today}
\pacs{83.80.Kn, 81.05.Rm, 82.35.Lr}

\maketitle

Microgels are microscopic polymer gels, swollen in a solvent, usually water. Due to their low internal polymer concentration, they are both mechanically soft and respond to changes in their physico-chemical environment such as solvent quality, pH, ionic concentration, and temperature. This sensitivity to external triggers is exploited in a range of industrial and biomedical applications, where the controlled swelling or deswelling of these systems can, for instance, be used for the controlled release of drugs and the manufacture of smart, responsive materials~\citep{Kesselman:Small:2012}.

Microgel particles or other soft objects such as biological cells generally adapt their volume in response to an applied osmotic pressure. This effect is often employed as a means to characterize their bulk elastic properties by measuring the pressure-dependent compression~\citep{Bastide:Macromolecules:1981,LI:1990fr,LietorSantos:PhysicalReviewE:2011,BonnetGonnet:Langmuir:1994,Polacheck:LabChip:2013}. However, kinetics associated with the equilibration process itself, which may potentially involve short and long time scales, is generally ignored. One reason for this is that the compression of these particles is generally much faster than the experimental time scale required to bring about a well-defined change in the applied osmotic pressure.
Hence, it is difficult to quantify the kinetics of the response of microgel particles to an osmotic shock in direct experiments. However, experimental access to this information would yield valuable insight into the viscoelastic properties of these materials that cannot be obtained by only studying the particle properties under (quasi-) equilibrium conditions.


In this article, we study these effects experimentally on a model system of microgel particles. To apply a rapid and well-controlled osmotic shock, we use dedicated microfluidic devices that enable us to subject microgel particles to rapid changes in osmotic pressure, while simultaneously visualizing the time-dependent changes in particle size by video microscopy. Surprisingly, we observe a response non-monotonic in time, where an initial rapid compression is followed by a slow reswelling of the particles after which they approximately regain their original size. The former we attribute to a poroelastic compression under constant external pressure, while the latter must be due to the penetration of the osmolyte into the microgel network, which in turn leads to a reduction of the osmotic pressure difference between the inside and outside of the microgel particles.

\begin{figure}[htb] \centering
\includegraphics[width=\linewidth]{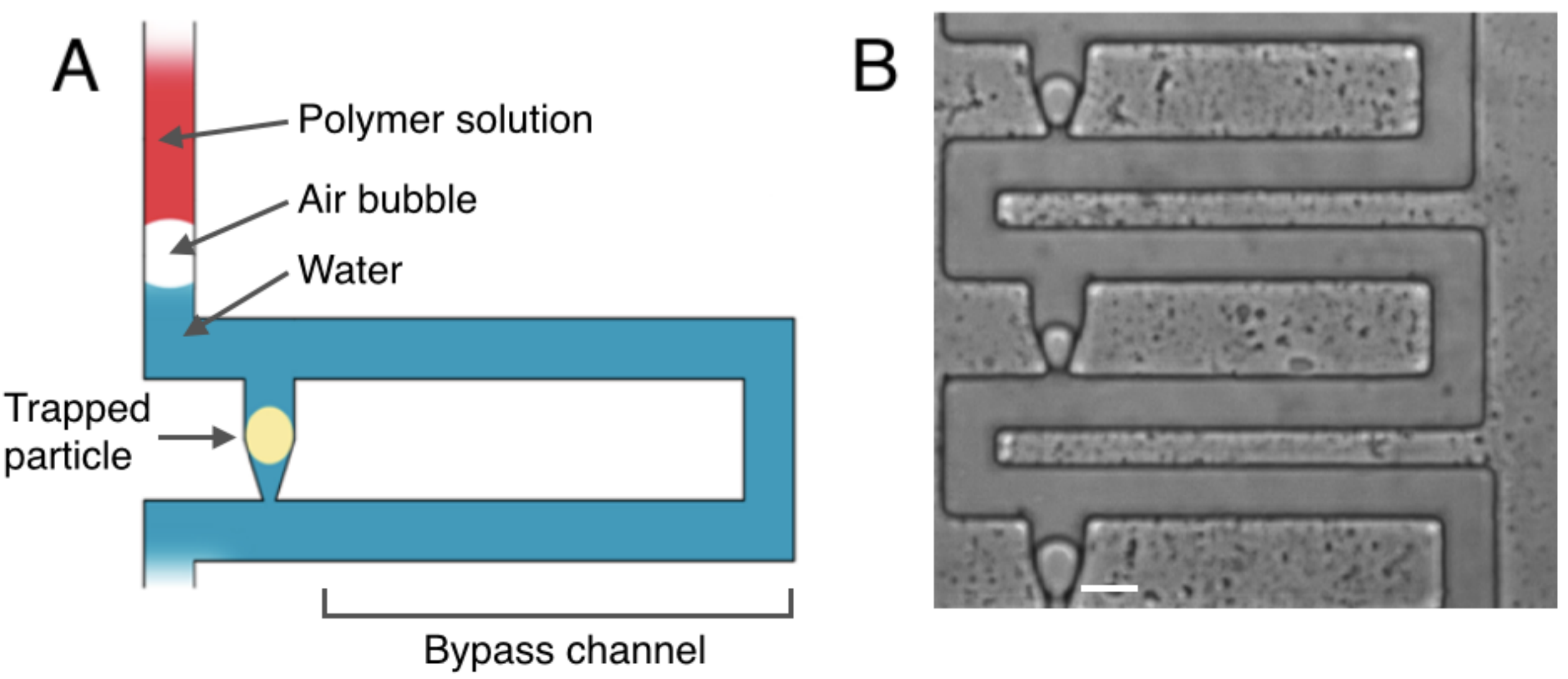}
\caption{Microfluidic particle trapping device for osmotic shock measurements. (A) Schematic of a single trap. (B) Typical device, showing trapped particles; scale bar is $40\ \mu\mathrm{m}$.}
\label{fig1}
\end{figure}

We perform our experiments on a model system of poly-acrylamide (PAA) microgel particles, synthesized by polymerization of aqueous drops in a water-in-oil emulsion. The aqueous phase contains all the reagents for the polymerization of PAA, 10 wt\% of acrylamide monomers, 12 wt\% of sodium chloride, as well as 0.5 wt\% of the cross-linker BIS-acrylamide. To obtain softer particles we also synthesize particles with a lower cross-linker concentration of 0.1 wt\% BIS-acrylamide. We then allow the polymerization reaction to take place inside the aqueous droplets by keeping the emulsion at a temperature of 65 $^\circ$C for $\sim$10 hours, resulting in the formation of a cross-linked polymer network. The resulting poly-acrylamide particles are then subjected to a series of washing steps, where the oil phase and any remaining unreacted monomers are removed from the solution by centrifugation, removal of supernatant, and dilution with de-ionized water (MilliQ, resistivity $\sigma > 18 \ \mathrm{M}\Omega\mathrm{cm}$).

Our microfluidic devices are fabricated from poly-dimethylsiloxane (PDMS) using standard soft lithography techniques~\citep{Xia:AngewChemIntEdit:1998}. They incorporate microfluidic particle traps similar to those described by Tan et al.~\citep{tan2007trap}, shown schematically in Fig.~\ref{fig1}~A. When particles initially enter the device, they flow into empty traps because the fluid resistivity of the trap is lower than that of the bypass channel. After a particle is caught in a trap, the situation is reversed and subsequent particles pass through the bypass channel. In a typical osmotic shock experiment, we first flow water past the trapped particles, followed by an aqueous solution of higher osmotic pressure, separated by an air bubble to avoid any diffusion- or convection-induced smoothing of the interface between the two fluids.

\begin{figure}[htbp]
\centering
\includegraphics[keepaspectratio,width=260pt,height=0.75\textheight]{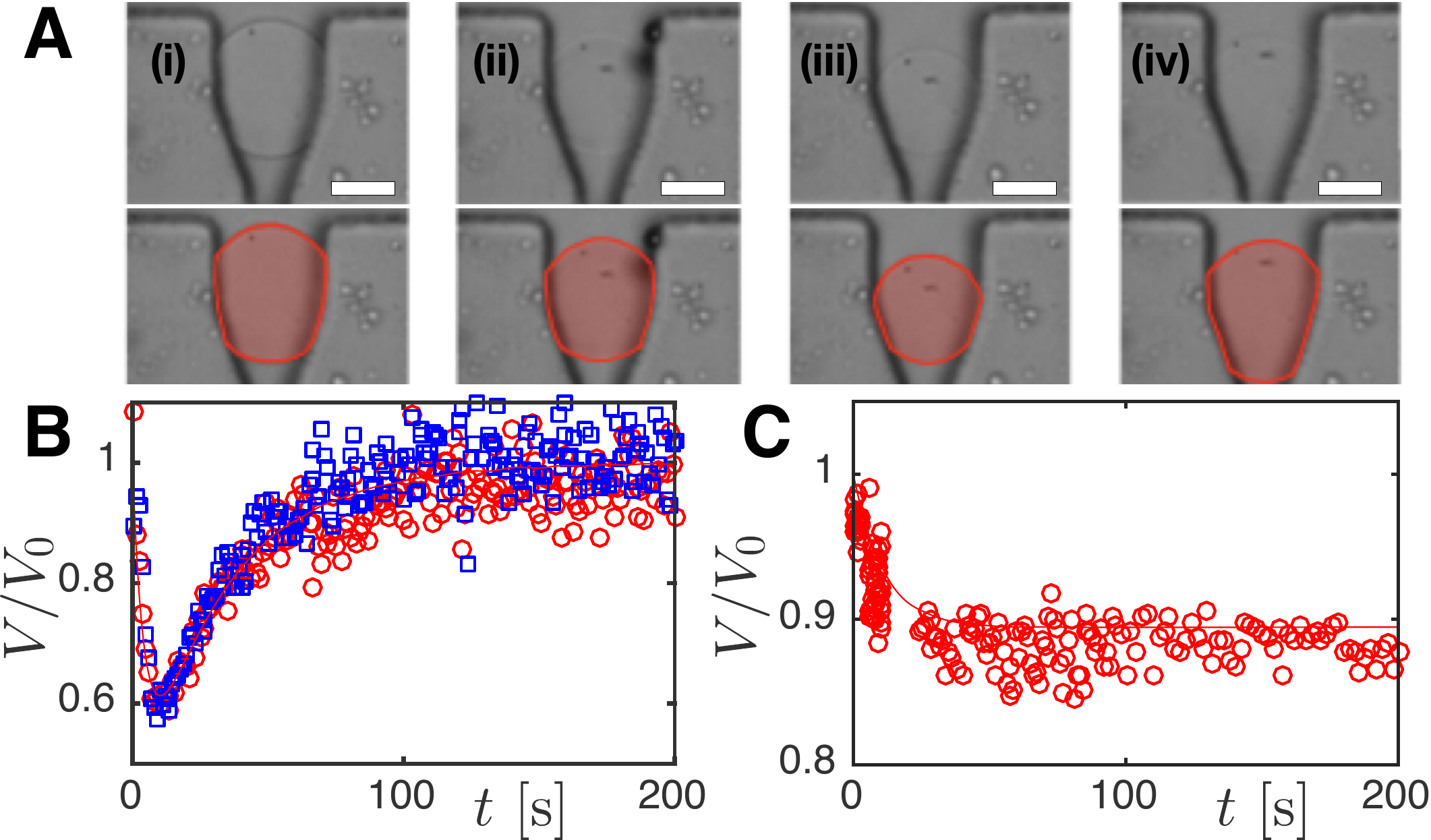}
\caption{Typical osmotic shock experiment. (A): Trapped PAA microgel particles, before the shock (i) at $t \approx 4 $ s (ii), near the minimum volume at $t \approx 10 \ \mathrm{s}$ (iii), and at $t \approx 100 \ \mathrm{s}$ (iv). Scale bar: 20 $\mu$m. Lower row: particle outlines highlighted for clarity. (B): $V(t)/V_0$ for flow rates 500 $\mu\mathrm{L/h}$ (circles) and 1000 $\mu\mathrm{L/h}$ (squares) following a shock to $\Pi_\mathrm{ext}=29.4$~kPa. (C): $V(t)/V_0$ for shock with high molecular weight osmolyte ($10^6$ g/mol PEO, $\Pi_\mathrm{ext}=1.5$~kPa).}
\label{fig2}
\end{figure}

During the entire process, we observe the particle of interest under a microscope and record the time-dependent changes in particle size by video microscopy. Representative frames are shown in Fig.~\ref{fig2}~A for a typical experiment of microgel particles ($0.5\ \mathrm{wt\%}$ cross-linker) under an osmotic shock from zero to $29.4\,\mathrm{kPa}$ using dextran ($M_\mathrm{r}=70~\mathrm{kg/mol}$, Sigma-Aldrich) as an osmolyte. In the bottom row of Fig.~\ref{fig2}~A we highlight the particle outlines for clarity.
For each frame of the corresponding movies, we estimate the particle volume $V(t)$ using digital image analysis (see SM).

We find that the particle responds initially by shrinking relatively swiftly, after which it slowly reswells to approximately its original size, as shown in Fig.~\ref{fig2}~B. The volume of the particle immediately after it is exposed to the high osmotic pressure remains unchanged, showing that the transfer of the bubble, as well as the exchange to a fluid of higher viscosity, has no significant impact on the particle volume. Hence, the subsequent compression of the particle must be caused solely by the increase in osmotic pressure. We test this by performing experiments at different flow rates, also shown in Fig.~\ref{fig2}~B, where we compare the scaled volume $V(t)/V_0$ as a function of time, $t$, for flow rates of 500 and 1000 $\mu$L\slash h, respectively.
Indeed, the data superimpose within experimental error.

We therefore perform all subsequent experiments at a fixed flow rate of 1000 $\mu$L\slash h, and vary only the osmotic pressure, as well as the cross-linking density of the microgel particles. The latter should influence the elastic response of the particles as well as the ease of transport of osmolyte into the particles. Through the former, we control the level of external stress applied to the particle. 

We hypothesize that the physical origin of the observed reswelling is governed by the penetration of osmolyte into the network of the microgel. A simple way to test this hypothesis is to use an osmolyte that cannot readily penetrate the polymer network due to its size.
Indeed, as shown in Fig.~\ref{fig2}~C, using a high molecular weight polyethylene oxide (PEO) solution ($M_w = 10^6\ \mathrm{g/mol}$ at a concentration of $2 \ \mathrm{wt}\%$, corresponding to $\Pi_\mathrm{ext} \approx 1.5$ kPa) to induce the osmotic shock, we observe only a rapid compression without any signs of reswelling, confirming our hypothesis.

\begin{figure}[htbp]
\centering
\includegraphics[width=\columnwidth]{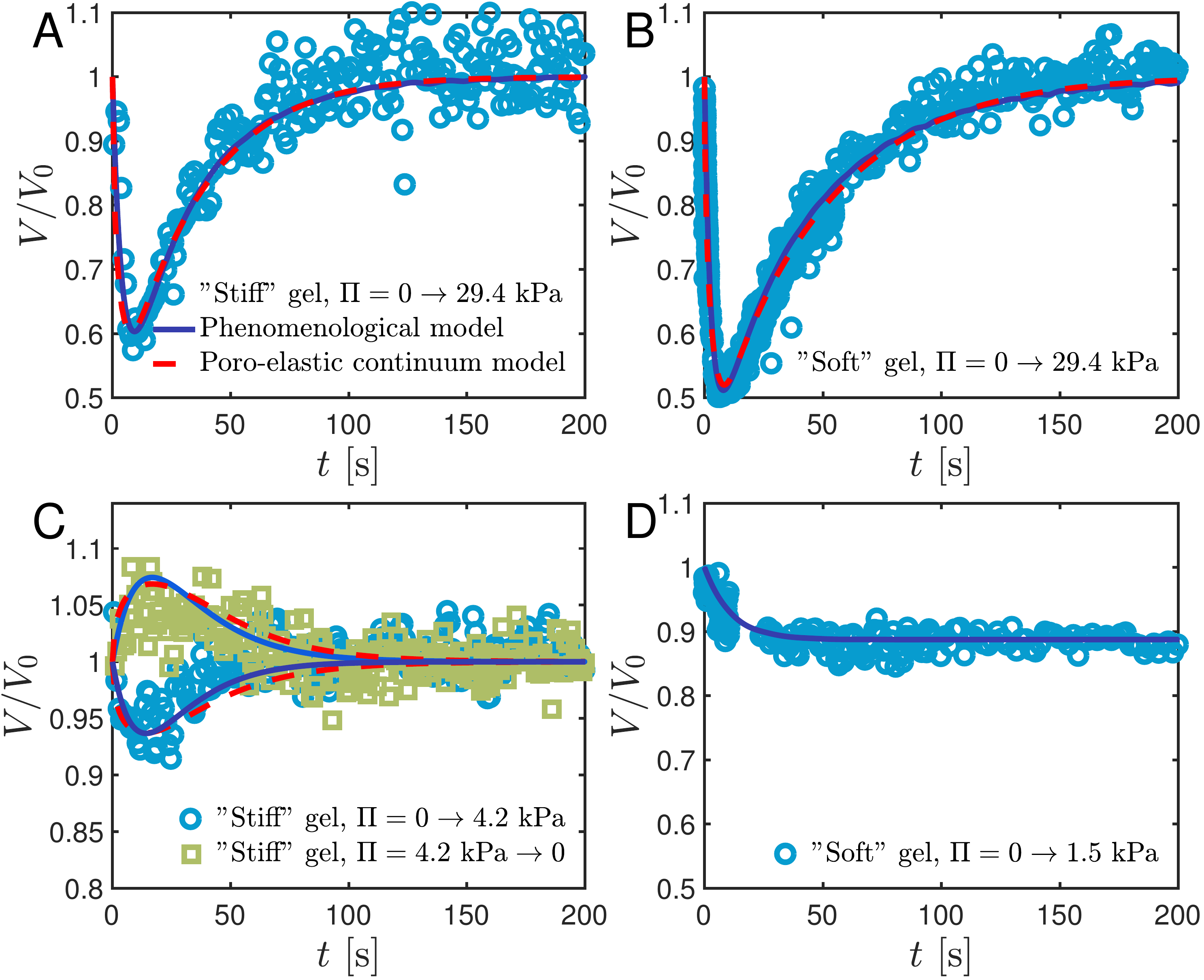}
\caption{
Experimental data and model fits for different experimental conditions, as indicated in each sub-figure. Osmolyte: 70~kg/mol dextran for (A-C) and $10^6$~g/mol PEO (D).
Fit parameters are collected in Table I.}
\label{fig3}
\end{figure}

Figure~\ref{fig3} summarizes our main experimental findings. Soft gels are compressed to a larger degree than stiff ones and take a longer time to reswell (Fig.~\ref{fig3}~A and B), exposing a microgel particle to a larger osmotic pressure induces a larger deformation (Fig.~\ref{fig3}~A and C). 
A microgel particle compressed and reswollen in a polymer solution and subsequently exposed to pure water, initially swells by solvent uptake to deswell at later times, where it again approximately recovers to its original size (Fig.~\ref{fig3}~C). In this process the polymer diffuses out of the microgel into the pure solvent. If the shock is performed with a high molecular weight osmolyte that cannot enter the pores (or does so very slowly), then, within the time scale of the experiment, no reswelling takes place (Fig.~\ref{fig3}~D).

To rationalize our findings we put forward a model that, despite its simplicity, captures the essential physics of the response of microgel particles to an osmotic shock. We assume that an equilibrated particle of initial radius $R_0$ is at time $t=0$ instantaneously exposed to a constant osmotic pressure $\Pi_\mathrm{ext}$, exerted by an osmolyte present in the surrounding fluid. Let the concentration of osmolyte at $t=0$ be uniform outside of the microgel particle and zero inside. As time progresses, osmolyte diffuses into the particle and the initially step-function-like concentration profile relaxes.

Plausibly the  diffusion of the osmolyte inside the particle is much slower than that outside the particle due to the presence of the polymer network. Moreover, in our osmotic shock experiments we are continuously flowing fresh osmolyte solution along the particle surface. Hence, we expect the osmolyte concentration and chemical potential at the surface to remain effectively constant. If this is true, we only need to consider the slow diffusion process within the microgel particle itself.

Because diffusion of osmolyte into the microgel is slow, the microgel responds elastically to the osmotic pressure difference caused by different concentrations of osmolyte in- and outside the microgel. If we rely on simple relaxational dynamics~\citep{Goldenfeld:1992wv}, then a sensible ansatz for a dynamical equation describing the time evolution of the radius $R(t)$ of the particle would be $\partial R/\partial t = - \Gamma \partial \Psi / \partial R$, where $\Gamma$ is a kinetic coefficient, $-\partial \Psi / \partial R$ is a generalized force, and $\Psi$ an appropriate free energy. The kinetic coefficient describes how easily the fluid is squeezed out of the microgel if it is under a compressive force.

The free energy $\Psi$ describes the mechanical and chemical work done during a volume change, and is a function of the elastic properties of the gel network. We simplify our description further by neglecting the spatial distribution of osmolyte within the particle, accounting only for the overall concentration within that particle. We further assume that the network consists of $m$ cross-linked subchains, and that it can be described as a uniform phantom network, which does not interact with the osmolyte~\citep{GrosbertKhokhlov}. This means that in equilibrium the internal and external concentration of the osmolyte will be equal. Considering that the volumes of our microgel particles are virtually indistinguishable before and after the osmotic shock experiment, this seems a reasonable assumption.

The mean density of the osmolyte in the microgel, $\rho = \frac{3N}{4 \pi R^3}$, depends on the number of osmolyte molecules in the microgel, $N$, and its radius $R$, if presumed spherical.
Our model free energy of the network including the absorbed osmolyte can now be written as $\Psi  = \frac{3}{2} m k_\mathrm{B} T \left( (R/R_0)^2 + (R_0/R)^2\right) + N k_\mathrm{B} T \left( \log(\rho \upsilon)-1 \right) - N \mu + \Pi_\mathrm{ext} 4\pi R^3/3$,
where $\upsilon$ is a microscopic volume scale and $\mu$ the chemical potential of the osmolyte that is set by its concentration in the fluid. The first two terms describe the ideal elastic behavior of the gel, and the third accounts for mixing and translation of the osmolyte within the microgel. The last two terms appear on account of the host fluid acting as both an osmolyte \emph{and} an osmotic pressure reservoir. If the background fluid behaves like an ideal solution, van't Hoffs law, $\Pi_\mathrm{ext}= \rho k_\mathrm{B} T$, applies and we can directly express the chemical potential in terms of its osmotic pressure, $\mu = k_\mathrm{B} T \log(\Pi_{\mathrm{ext}} \upsilon/k_\mathrm{B} T)$.
This produces the following dynamical equation for the ratio $\alpha (t) = R(t)/R_0$ between the radius at time $t$ and at time zero, $ \partial \alpha / \partial t = - 3 \Gamma_\alpha \left[ \alpha - \alpha^{-3} - P \left( f \alpha^{-1} -  \alpha^2 \right) \right]\ ,
$
where $P \equiv \Pi_\mathrm{ext} /  K$ is a dimensionless osmotic pressure scaled to the bulk compressive modulus $K = 3 m k_\mathrm{B}T  / 4 \pi R_0^3$ of the network, and $\Gamma_\alpha \equiv \Gamma k_\mathrm{B} T m / R_0^2$ is a phenomenological relaxation rate. Below we express this rate in terms of the elastic modulus and permeability of the network and the viscosity of the solvent. The appropriate initial condition is $\alpha(0)=1$.

To account for diffusion of the osmolyte from the fluid into the microgel, we invoke the diffusion equation in integral form, $\partial N/\partial t = D \oint d^2 S \cdot \left[ \rho \nabla \mu / k_\mathrm{B}T \right]$ across the interface, where $D$ is the diffusivity of the osmolyte within the gel. Within our coarse-grained model prescription in which the osmolyte behaves ideally and where we treat the concentrations in- and outside the microgel particle as uniform but different, this becomes $\partial N / \partial t = D R^{-2} (N_\infty-N)$ with $R(t)$ the radius of the microgel and $N_\infty$ the equilibrium value of the number of osmolyte particles; all constants of proportionality are absorbed in the diffusivity.

If we define $f(t) = N(t)/N_\infty$ as the fraction of the total equilibrium amount of osmolyte in the microgel at time $t$, we obtain $ \partial f / \partial t = - \Gamma_f \alpha^{-2} (f-1)$,
where we have introduced the kinetic parameter $\Gamma_f \equiv D / R_0^2$ that sets the time scale for the solute to enter the microgel by diffusion. An obvious initial condition is $f(0)=0$. The ratio of the rates $\gamma = \Gamma_f/3\Gamma_\alpha$ determines to what extent the microgel particle can be compressed if exposed to an instantaneous osmotic stress and also determines the time scale over which reswelling occurs.

The reverse case of exposing an osmolyte-saturated microgel particle to a solvent devoid of any osmolyte, produces slightly different differential equations that within our treatment read $\partial \alpha / \partial t = - 3\Gamma_\alpha [\alpha - \alpha^{-3} - fP]$ and $\partial f / \partial t = - \Gamma_f \alpha^{-2} f$, where $P$ has the same meaning as before, describing the osmotic pressure of the solution before it is replaced by pure solvent, and $f(t) = N(t)/N(0)$ is the fraction of osmolyte depleted from the microgel particle. Initial conditions are $\alpha(0)=1$ and $f(0)=1$. In this case the particle swells immediately after immersion into pure solvent because of the osmotic pressure exerted by the osmolyte trapped within the microgel.

Our time evolution equations are highly non-linear and we have not been able to solve them analytically.
The short-time solution to the linearized equations is bi-exponential, but in the limit of slow polymer penetration it is dominated by one of the relaxation rates, equal to $3\Gamma_\alpha (12+3P)$. The long-time response is in that case also bi-exponential with the slower rate being $\Gamma_f$. This shows that the relevant time scales are functions of the diffusivity of the osmolyte  (through $\Gamma_f$), the cross-linking density and the permeability of the microgel (through $\Gamma_\alpha$ and $P$), the osmotic pressure of the solution (through $P$), as well as the gel particle size (through $\Gamma_f$, $\Gamma_\alpha$ and $P$).

We have also developed a more detailed model based on conventional poroelastic theory~\citep{MacMinn:2016hv}, in which we take into account the local force balance between the osmotic pressure of the dissolved polymers and the mechanical response of the network.
To be able to compare this more elaborate model to our phenomenological model, we assume the dextran inside the hydrogel to be unhindered by the PAA network, i.e., we again consider the PAA network to be a phantom network. The full geometric non-linearity of large deformations we capture by assuming Hencky elasticity for the effective stress~\citep{MacMinn:2016hv}. Interaction between the PAA network and the water we describe making use of Darcy's law~\citep{Tartakovsky:PhysicalReviewLetters:2008}. More details can be found in the SM and in a follow-up article that we intend to submit for publication soon.

We evaluate numerically the governing equations for both models, optimizing the various model parameters against the experimental data.
As shown in Fig.~\ref{fig3}, this yields a surprisingly good agreement between both models and the experimentally observed compression\slash reswelling and swelling\slash recompression curves. In Table~\ref{table1} we collect the fitted model parameters for the experiments shown in Fig.~\ref{fig3}. We can directly relate the phenomenological model parameters obtained from these fits to more intuitive material properties, such as the permeability of the gel network to the solvent water, $\kappa$, the bulk compressive modulus of the network, $K$, and the diffusion constant of the osmolyte in the network, $D$.\\
The compressive modulus we obtain directly from the fitted value of $P \equiv \Pi_\mathrm{ext} / K$, and the diffusion coefficient of the osmolyte in the network from $D=\Gamma_{f} R_0^2/\pi^2$. To find the permeability of the network, we identify our initial (de)swelling rate with the slowest relaxation in the swelling process of a microgel particle from the model of Tanaka and Fillmore~\citep{tanaka1979kinetics} as $3 \Gamma_\alpha(12 + 3P) \approx \pi^2 K \kappa / R_0^2 \eta$, where $\eta\approx 10^{-3}$~Pa\,s is the viscosity of water.

The obtained compressive bulk modulus of 13--14 kPa for the ``stiff'' microgel particles is in good agreement with the value of $K \approx 13 \pm 5 \ \textrm{kPa}$ that we obtained from an independent Capillary Micromechanics measurement~\citep{Wyss:SoftMatter:2010,Guo:MacromolMaterEng:} (see SM). Similarly, the values we obtain for the gel's permeability ($\kappa \approx 0.1\textendash 0.2\ \mathrm{nm}^2$) are consistent with values found in literature for similar hydrogels~\citep{Tokita:TheJournalOfChemicalPhysics:1991a}.
Further, while we did not find literature data for the diffusion coefficient of dextran in PAA gels, the obtained diffusion coefficient of $D \approx 1\ \mu\mathrm{m}^2/\mathrm{s}$ for 70~kg/mol dextran within the PAA networks is indeed much lower than in water, where $D \approx 30\ \mu\mathrm{m}^2/\mathrm{s}$~\cite{N:BiophysicalJournal:1998}.

\begin{table}[h!]
\centering
\begin{tabular}{lccccc}
Data in Fig. :                                                     & \textbf{3A}   & \textbf{3B}   & \textbf{3C}   & \textbf{3C}   & \textbf{3D}       \\
\hline
$\Pi_\mathrm{ext}$ [kPa]                                                   & 29.4 & 29.4 & +4.2  & -4.2 & 1.5  \\
\multicolumn{6}{l}{\ \emph{Parameters phenomenological model:}}                                      \\
$K$ [kPa]                                                     & 13   & 10   & 13   & 13   & 10   \\
$t_\mathrm{fast} = \frac{\pi^2}{3 \Gamma_\alpha (12+3P)}$ [s] & 32   & 19   & 61   & 72   & 43   \\
$t_{\mathrm{slow}} = {\pi^2}/{\Gamma_f}$ [s]               & 395  & 493  & 132  & 132  & --   \\
$D$ [$\mu\mathrm{m}^2$/s]                                     & 1.0 & 0.8 & 3.0 & 3.0 & --   \\
$\kappa$ [nm$^2$]                                             & 0.10 & 0.21 & 0.05 & 0.04 & 0.09 \\
\multicolumn{6}{l}{\ \emph{Parameters poro-elastic model:}}                                          \\
$K$ [kPa]                                                     & 14   & 11   & 14   & 14   & --   \\
$t_\mathrm{gel}$ [s]                                          & 60   & 53   & 60   & 60   & --   \\
$t_\mathrm{pol}$ [s]                                          & 240  & 350  & 240  & 240  & --   \\
$D$ [$\mu\mathrm{m}^2$/s]                                     & 1.7  & 1.1  & 1.7  & 1.7  & --   \\
$\kappa$ [nm$^2$]                                             & 0.16 & 0.23 & 0.16 & 0.16 & --
\end{tabular}
\caption{Parameters and extracted material properties for curve fits shown in Fig.~\ref{fig3}. Input parameters are the particle radius, $R\approx 20 \ \mu\mathrm{m}$, and the presumed Poisson ratio, $\nu=0.48$~\cite{takigawa1996poisson}.}
\label{table1}
\end{table}

In summary, our method enables direct experimental access to three key physical properties of porous soft objects: their elastic bulk modulus, their permeability to an aqueous background liquid, and the mobility of osmolyte macromolecules within the microgel network.
While at the macroscopic scale measurement of each of these properties requires separate, dedicated techniques and instruments, using our microfluidic approach they become readily accessible in one simple experiment. We expect our approach to be relevant to applications and materials where the properties of soft, compressible objects are of key importance, such as in ``smart'', responsive materials.

\newpage

\newpage

\onecolumngrid
\appendix

\section{Supplemental Material (SM) for article "Compression and reswelling of microgel particles after an osmotic shock"}

\subsection{I. Supplementary video}  
\label{supplementalvideo}

The video ``Supplementary\_video1.mp4'' shows a typical experiment, as recorded with video microscopy.\\
\emph{Experimental conditions}: A poly-acrylamide (PAA) particle of the ``soft'' type (cross-linker concentration of 0.1 wt\% BIS-acrylamide) is exposed to an osmotic shock from deionized water to a 13 wt\% dextran solution ($M_\mathrm{w}=70~\mathrm{kg/mol}$, osmotic pressure $\Pi \approx 29.4\ \mathrm{kPa}$). The two fluids are separated by an air bubble, which in the movie appears at $t \approx 23 \ \mathrm{s}$ and fully disappears at $t \approx 48 \ \mathrm{s}$, at which point the PAA particle is surrounded by the high osmotic pressure solution. The particle rapidly shrinks, reaching its smallest size at around $t \approx 55 \ \mathrm{s}$ in the movie. Subsequently, the particle slowly re-swells to nearly its original size in the rest of the video.

\subsection{II. Mechanics of microgel particles: Capillary Micromechanics}
\label{mechanicsofmicrogelparticles:capillarymicromechanics}

As a comparison to our osmotic shock experiments, we quantify the mechanical properties of our hydrogel particles using the recently developed Capillary Micromechanics method~\citep{Guo:MacromolMaterEng:}. The results serve as a validation for our main measurements, performed in the microfluidic osmotic shock setup.\\
In Capillary Micromechanics, a dilute suspension of the particles of interest is flown into a tapered glass capillary of circular cross-section. As the tip of the capillary is smaller in diameter than the particles, the first particle that arrives near the tip remains trapped and subsequently blocks the further flow of fluid through the capillary. In this situation, the entire pressure difference applied across the capillary falls off across the trapped particle. The corresponding applied external stress must match the internal elastic stress within the particle, which is a function of the particle's deformation and the elastic moduli of the particle. Thus, quantifying the deformation of the particle enables us to directly extract its elastic response. In particular for the case of the compressive (bulk) modulus $K$, we quantify the change in volume as a function of the characteristic bulk stress $\sigma_\mathrm{compr.} \approx \left[2 \cdot p_\mathrm{wall} + p \right] /3$ applied to the particle, where $p_\mathrm{wall}$ is the pressure exerted on the particle at the area of contact between the particle and the wall of the capillary, and $p$ is the pressure drop applied across the capillary.~\citep{Guo:MacromolMaterEng:} 

\subsubsection{A. `Stiff' particles}
\label{stiffparticles}

\begin{figure}[htb] \centering
\includegraphics[width=0.6\linewidth]{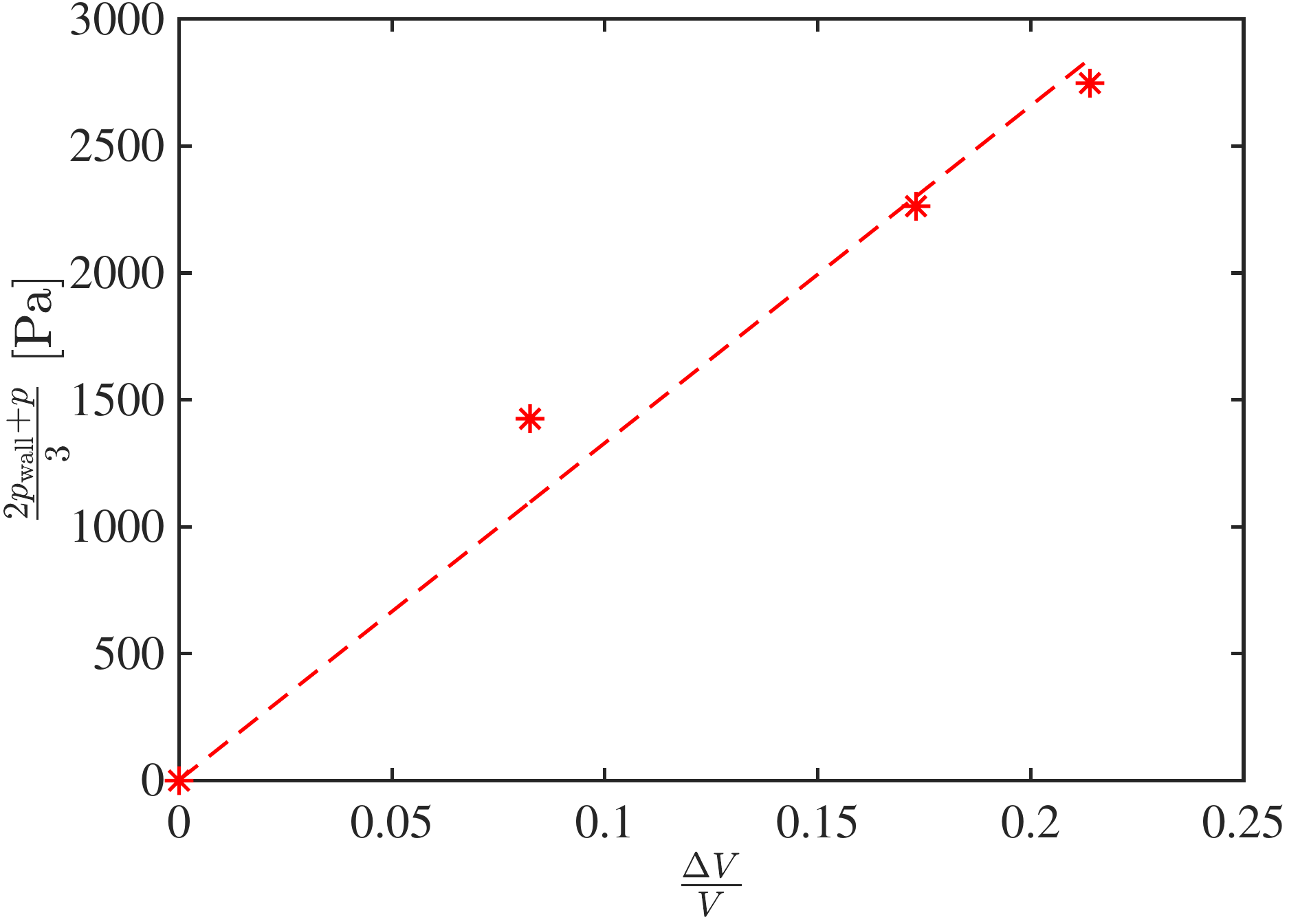}
\caption{Capillary Micromechanics measurement of the bulk (compressive) modulus $K$ for our 'stiff' particles (0.5 wt\% BIS-acrylamide): The characteristic stress $(2 p_\mathrm{wall} + p ) /3$ as a function of the volumetric strain $\Delta V/V$. The slope of this curve corresponds to the bulk modulus $K$. A linear fit, indicated by the dashed line, yields $K \approx 13\ \mathrm{kPa}$}
\label{capmicro}
\end{figure}

Results from a measurement on our `stiff' particles (cross-linker concentration of 0.5 wt\% BIS-acrylamide) is shown in Fig.\ref{capmicro}. The characteristic stress for compression, $(2 p_\mathrm{wall} + p) /3$, is plotted as a function of the volumetric strain $\Delta V/V$. A linear fit to the data, shown as the dashed line, yields $K \approx 13\ \pm 5\ \mathrm{kPa}$.

\subsubsection{B. `Soft' Particles}
\label{softparticles}

For our `soft' particles (cross-linker concentration of 0.1 wt\% BIS-acrylamide) the measurements proved difficult, as -- due to their low internal polymer concentration in the swollen state -- the particles are difficult to visualize and exhibit lower elastic moduli than the `stiff' hydrogel particles. As a result, we did not achieve valid measurements on these particles, instead only being able to identify that particles were flowing through the constriction at pressures where the `stiff' particles were trapped, showing that their elastic response is indeed weaker than that of our `stiff' particles.

\subsection{III. Osmotic pressure of polymer solutions}
\label{osmoticpressureofpolymersolutions}

\subsubsection{A. Dextran solutions}
\label{dextransolutions}

In our main experiments, we apply an external osmotic pressure on our particles using a solution of dextran with a molecular weight of 70 kg\slash mol ($M_\mathrm{r}=70~\mathrm{kg/mol}$, Sigma-Aldrich, cat. nr. 31390).
For this grade of dextran the osmotic pressure as a function of concentration has previously been measured in detail by Bonnet-Gonnet et al.~\citep{BonnetGonnet:Langmuir:1994}
Their experiments showed that the osmotic pressure within a range of concentrations from 0.1\% to 15\% is well described by a polynomial expansion, as $\Pi(\tilde{c}) \approx \left[ 286 {\tilde{c}} + 87 {\tilde{c}}^2 + 5 {\tilde{c}}^3 \right] \ \mathrm{Pa} $, where ${\tilde{c}} = c / (1 \mathrm{wt}\%)$ is the polymer concentration non-dimensionalized by 1 wt\%.

We here use this approximation to determine the osmotic pressure of our solutions from the polymer concentration. For the dextran concentrations used in our experiments, $c=5\ $wt\% and $c=13\ $wt\%, this yields 4.2 kPa and 29.4 kPa, respectively. 

\subsubsection{B. Poly-ethylene oxide (PEO) solutions}
\label{poly-ethyleneoxidepeosolutions}

To apply an osmotic pressure in the absence of osmolyte penetration into the poly-acrylamide particles, we use high molecular weight poly-ethylene oxide (PEO) solutions ($M_\mathrm{w}=1000\ \mathrm{kg}/\mathrm{mol}$, Sigma Aldrich) instead of the 70kD dextran solutions used in our main experiments. We measure the osmotic pressure of the PEO solutions as a function of concentration by dialysis against dextran solutions, for which the osmotic pressure has previously been measured in detail~\citep{BonnetGonnet:Langmuir:1994}.\\
We enclose $1\,\mathrm{wt}\%$ PEO solutions into dialysis bags and place them into baths of dextran solutions with concentrations ranging from $c=1\ \mathrm{wt}\%$ to $c=4\ \mathrm{wt}\%$, corresponding to osmotic pressures between $\approx 0.38$ to $\approx 2.8\ \mathrm{kPa}$. These baths are then allowed to equilibrate for a period of 1 week; after this period we assume that the osmotic pressure of PEO inside the bag matches that of dextran in the bath outside. Depending on whether the initial osmotic pressure in the PEO solution is larger or smaller than that in the surrounding dextran solution, the concentration within the dialysis bag will have increased or decreased after equilibration, respectively. In our further analysis, we neglect changes in concentration in the dextran bath around the dialysis bag, as the volume of the dextran solution far exceeds the volume of the sample in the dialysis bag (ratio $>$ 50:1).

\begin{figure}[htb] \centering
\includegraphics[width=0.6\linewidth]{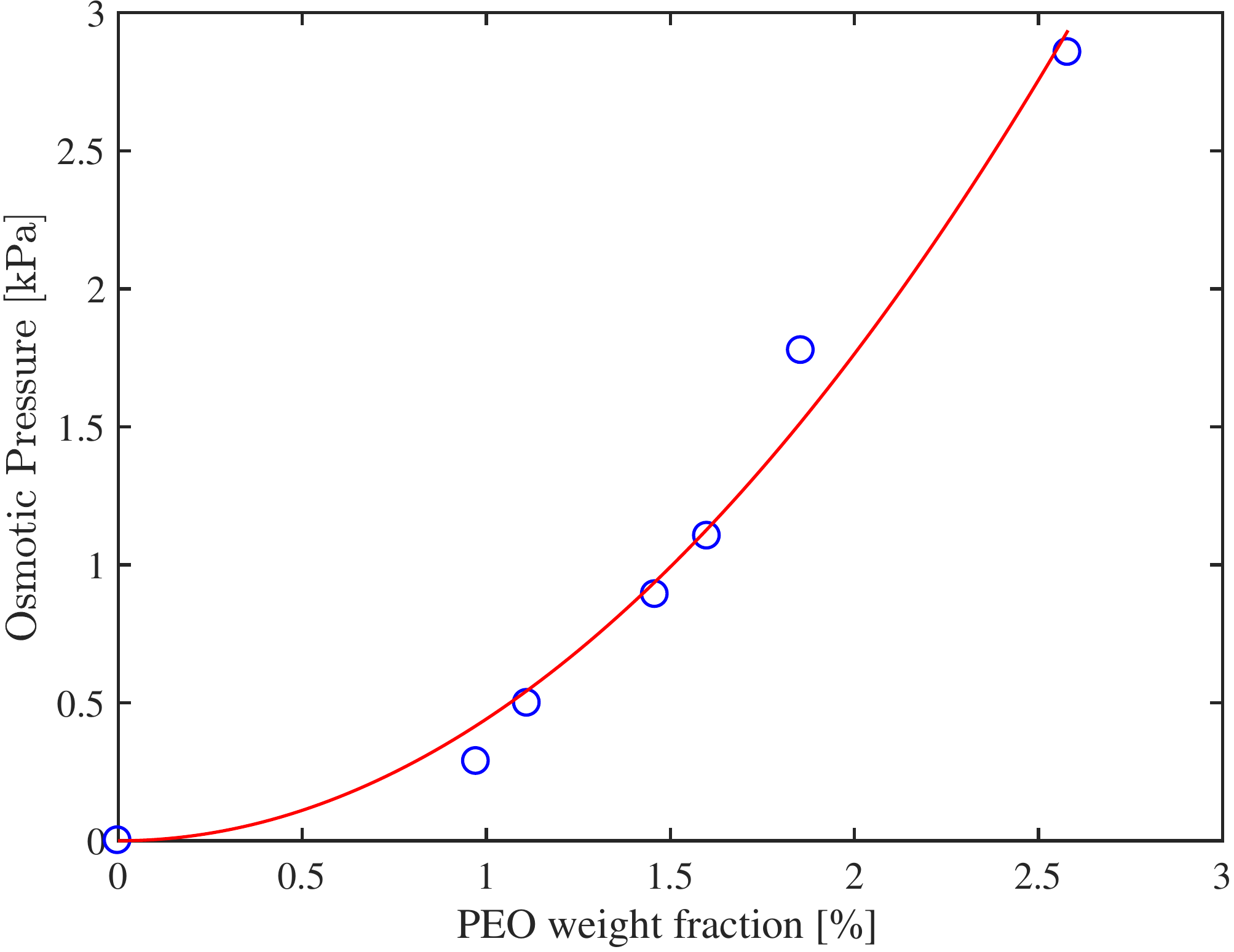}
\caption{Osmotic pressure as a function of concentration for PEO solutions, determined from dialysis against dextran solutions.}
\label{PEO_P_osm}
\end{figure}

To determine the concentrations of the equilibrated PEO solutions, we extract the solutions from the dialysis bags and determine their weight both immediately after extraction and after thorough overnight drying on a hotplate, respectively. The PEO weight concentration is then taken as the ratio of the dry weight to the initial weight of the solution. The resulting data points for the different solutions are shown in Fig.\ref{PEO_P_osm} as blue circles. The red line is a second order polynomial fit to the data, meant to guide the eye. Results are in reasonable agreement with the osmotic shock experiments, which suggest an osmotic pressure of $\approx 1.5\,\mathrm{kPa}$ for a $2\,wt\%$ PEO solution.

\subsection{IV. Image analysis for estimating particle volumes}
\label{imageanalysisforestimatingparticlevolumes}

\begin{figure}[htb!] \centering
\includegraphics[width=0.5\linewidth]{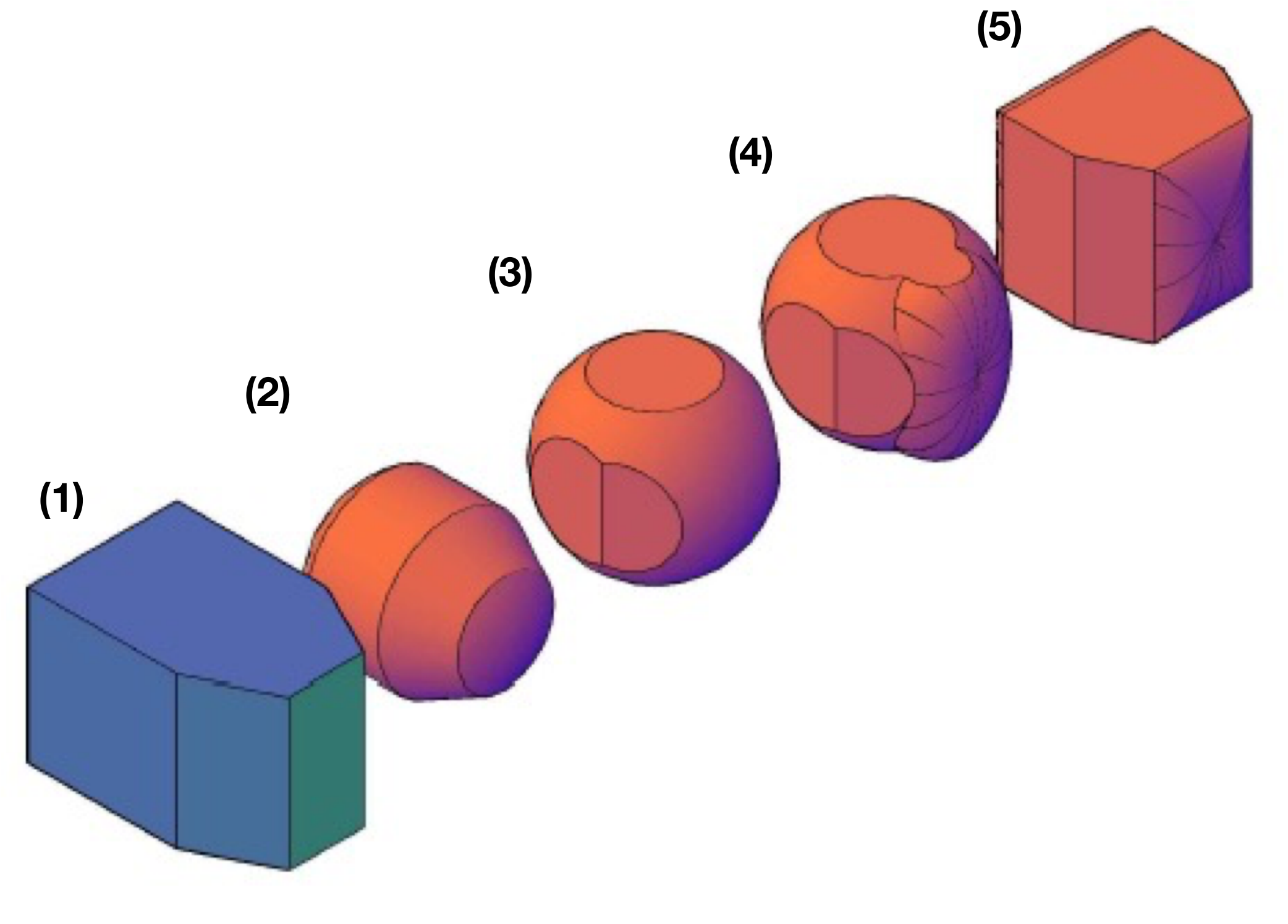}
\caption{
Possible approximations of 3D particle shape. \textbf{(1)} Shape of the microfluidic trap \textbf{(2)}  Lower limit: Volume obtained by rotating the 2D particle shape observed in the microscope around the central axis along the middle of the trap. \textbf{(3)} Volume obtained starting from a 3D ellipsoid that is deformed locally by the PDMS wall, otherwise keeping its exterior integrity where the wall is not in contact. \textbf{(4)} Volume obtained from the 3D ellipsoid that is deformed in the straight part of the trap only where it is in contact with the wall and where the particle maintains ellipsoidal cross-sections in the tapered part of the microfluidic trap. \textbf{(5)} Upper limit: Volume obtained by extruding the 2D particle shape across the full height of the channel.}  
\label{imageanalysis}
\end{figure}

For each osmotic shock experiment, we follow the time-development of the particle's shape and volume using video microscopy. We thereby obtain, for each frame of the recorded video, the two-dimensional projection of the particle shape, as seen from the top of the channel. To obtain an estimate of the three-dimensional particle shape, different approximations can be employed, as shown in Fig.\ref{imageanalysis}. We find that the 2D projection of the particle shape is well approximated by an ellipsoid that is locally deformed (as in a Hertz approximation) by the walls of the microfluidic channel. This is appropriately reflected in case (3) shown in Fig.\ref{imageanalysis}, where we assume the 3D shape to be a 3D ellipsoid that is locally deformed by the walls of the device. Comparison of volumes obtained using cases (2), (3) and (4), we find relatively small differences, varying by typically less than 20\%, even at the largest degrees of deformation probed in our experiment. Moreover, the general trend of a rapid volume decrease, followed by a slow re-swelling, is observed independent of what approximation for the 3D particle shape is taken. We therefore choose to use the approximation represented by case (3) in Fig.\ref{imageanalysis} to extract particle volumes from our osmotic shock experiments.\\
To do so, we first determine the two-dimensional shape of both the particle and the microfluidic by using a home-built MATLAB code employing edge detection methods to detect the particle's shape. The portions of the particle outline that are not in contact with the wall of the microfluidic trap are fitted to an ellipsoid, which, as mentioned above, agrees surprisingly well with the actual particle shape, even for significantly deformed particles. We then calculate the 3D shape and volume of the particle, based on approximation (3) shown in Fig.\ref{imageanalysis}.

\subsection{V. Poro-elastic model}
\label{poro-elasticmodel}

In this paragraph we briefly outline the ingredients of the poro-elastic continuum model we use to model the observed osmotic shock. The water phase and the polyacrylamide (PAA) network are both modeled as a continuum. The dissolved dextran polymers are considered to be a constituent of the water phase ~\citep{Klika:CriticalReviewsInSolidStateAndMaterials:2014}. 

The equations of motion are found by considering conservation of water and PAA volume and the number of dissolved polymers inside the hydrogel (see the book by Coussy~\citep{Coussy:Poromechanics:2004}, chapter 1). The diffusive flux of polymers is modeled assuming infinite dilution and no hinder of the PAA network, implying Fick's first law to apply with a constant diffusion coefficient.
The final equation of motion is given by the overall force balance of the two continua (see ~\citep{Coussy:Poromechanics:2004}, chapter 2). Stress inside the hydrogel is generated by the fluid pressure, the Terzaghi effective stress of the PAA network~\citep{MacMinn:2016hv}, and the osmotic pressure of the dissolved polymers. To find the effective stress of the network we consider Hencky elasticity as a constitutive relation. It captures the full geometric nonlinearity of large deformations as it is essentially a three dimensional version of the one-dimensional concept of true strain with a constant osmotic and shear modulus~\citep{MacMinn:2016hv}. As PAA forms a very ``rubber like'' network, we expect this approximation to hold. The osmotic pressure of the dissolved polymers is given by van 't Hoffs law, for we model the dissolved polymers as being infinitely diluted. Darcy's law accounts for interaction between the two continua and provides an expression for the fluid pressure~\citep{Tartakovsky:PhysicalReviewLetters:2008}. Tokita \& Tanaka ~\citep{Tokita:TheJournalOfChemicalPhysics:1991a} established the dependence of the permeability on the local concentration of PAA network. Finally, assuming radial symmetry of the hydrogel, we can express the velocity of the water continuum in that of the PAA network by combining the volume conservation equations of the two continua. 

The boundary conditions needed to solve the equations of motion are found from the following considerations. In the microfluidic setup, polymer solution flows continuously around the hydrogels. As the dynamics of the osmotic shock are independent of the flow speed of the solution, we assume a constant concentration of polymers at the boundary of the hydrogel. If the PAA network is permeable for the dissolved polymers the concentration of polymers inside and outside the hydrogel is equal at the boundary. At the center of the gel the concentration of polymers acquires a local minimum because the concentration of dissolved polymers is continuously differentiable at the center. The effective stress is zero at the boundary because we neglect the pressure exerted by the microfluidic device, and because the concentration of polymers inside and outside the hydrogel is equal at the boundary. At the center of the hydrogel the deformation of the PAA continuum should be zero because of symmetry. If the dissolved polymers can not penetrate into the hydrogel the concentration of polymers inside and outside the hydrogel is not equal at the boundary, implying the effective stress at the boundary to be equal to the osmotic pressure of the polymer solution.


\end{document}